\documentstyle[12pt,a4,amssymb,times]{article}

\def\a{\alpha}
\def\vf{\varphi}
\def\f{\phi}
\def\th{\theta}

\def\e{\epsilon}

\def\g{\gamma}
\def\G{\Gamma}
\def\O{\Omega}
\def\o{\omega}
\def\p{\pi}
\def\d{\delta}
\def\m{\mu}

\def\l{\lambda}
\def\z{\zeta}
\def\ca{{\cal A}}
\def\cb{{\cal B}}

\def\to{\rightarrow}
\def\con{{\rm const.}}
\newcommand{\be}{\begin{equation}}
\newcommand{\ee}{\end{equation}} 
\newcommand{\bea}{\begin{eqnarray}}
\newcommand{\eea}{\end{eqnarray}}

\begin{document}
\begin{titlepage}

\begin{flushright} 
December 1999
\end{flushright}

\bigskip
\bigskip
\bigskip
\bigskip

\begin{center}

{\bf{\Large Matrix Factorization for an $SO(2)$ Spinning Top\\
and Related Problems}}

\bigskip 

 A. Mikovi\'c \footnote{E-mail address: amikovic@ualg.pt. On leave of absence 
from Institute of Physics, P.O.Box 57, 11001 Belgrade, Yugoslavia}

\end{center}

\begin{center}

\footnotesize
	 \'Area Departamental de Matematica, UCEH, Universidade do Algarve,
         Campus de Gambelas, 8000 Faro, Portugal 
\end{center}

\normalsize 

\bigskip 
\bigskip
\begin{center}
			{\bf Abstract}
\end{center}	
We study the matrix factorization problem associated with an $SO(2)$
spinning top by using the algebro-geometric approach. We derive the
explicit expressions in terms of Riemann theta functions and discus some 
related problems including a
non-compact extension and the case when the Lax matrix contains higher-order 
powers of the spectral parameter. 

\noindent
\bigskip
\bigskip

\end{titlepage}
 \newpage

\section{Introduction}

Analytic factorization for matrix valued functions, or the 
matrix Riemann-Hilbert problem, appears in problems of
exactly integrable dynamical systems as well as in problems of elasticity and 
diffraction.

In the context of exactly integrable models, the
factorization problem for the R-matrix associated to an affine Lie algebra 
is the same as 
the matrix Riemann-Hilbert problem \cite{rsts}. In this case
exists a general formula for the solution in terms of Riemann theta 
functions \cite{rsts}, derived from algebro-geometric methods of finite-band
integration theory \cite{fbi}. This construction is based on the fact that
the time evolution linearizes on the Jacobian of the spectral curve, and 
consequently one can express the solution in terms of the Baker-Akheizer 
functions, which are eigen-vectors of the Lax matrix.
Various integrable dynamical systems can be
formulated via such R-matrix \cite{rsts}, and most typical are the
spinning tops. 

In this paper we derive the
explicit expressions for the case of an $SO(2)$ spinning top, which is a 
simple but a non-trivial example of the two-by-two matrix factorization 
problem. This is done in order to illustrate the general theta-function
formula from \cite{rsts}, and also
in order to be able to compare to alternative methods of matrix factorization.
The neccesity for alternative methods of factorization comes from the fact that
the theta-function formula becomes difficult to use in the case when the
spectral curve is non-hyperelliptic, which is mostly the case. Also there are
cases when only a non-canonical factorization is possible, for which the
theta-function formula is not valid. We illustrate this on the example of a
non-compact extension of the $SO(2)$ spinning top.

\section{Lax formulation and matrix Rieman-Hilbert problem}

Consider a rotator with a unit radius and angle $\f$ in a potential
$V(\f)=-a^2 \cos 2\f$, where $a$ is a real constant. The Hamiltonian is 
given by
\be
H = \frac12 l^2 - a^2 \cos 2\f \quad, \label{hrot}
\ee
where $l = \dot\f =d\f /dt$. As a one-dimensional Hamiltonian system, it is 
integrable,
and it is the $n=2$ case of an $n$-dimensional integrable spinning top
system given by the following $n$-by-$n$ Lax matrix \cite{rsts}
\be 
L = {\bf a}\l + {\bf l} + {\bf s} \l^{-1} = {\bf a}\l + 
{\bf l} +  R_{\f}^{-1} {\bf f} R_\f \l^{-1} 
\quad,\label{lm}
\ee
where $\l$ is a complex number called the spectral parameter, $\bf a$ and 
$\bf f$ are fixed symmetric 
matrices, ${\bf l}$ is antisymmetric and traceless matrix, while $R_\f$ is an 
$SO(n)$ matrix. Formula (\ref{lm}) gives a mapping from the phase space of the 
$n$-dimensional spinning top $T^* SO(n)= (R_\f,{\bf l})$ to a Hamiltonian 
orbit in
the affine $gl(n)$ algebra $\hat g$ defined as 
$$\hat g = \sum_{j\in {\bf Z}} g_j \l^j \quad,\quad g_j \in gl(n)\quad.$$
Without loss of generality one can 
take that
the $\bf a$ and $\bf f$ matrices are diagonal, and for simplicity we 
take ${\bf a}={\bf f}$ and
$tr {\bf a}=0$. The last condition implies $tr\, L =0$, and this means that 
we are restricting $\hat g$ to the affine $sl(n)$ sub-algebra.

The Hamiltonian (\ref{hrot}) is the $n=2$ case of the 
$SO(n)$ Hamiltonian
\be
H = -\frac12 Res\,tr(\l^{-1} L^2) = -\frac12 tr{\bf l}^2 - 
tr ({\bf a}R_\f^{-1}{\bf f}R_\f)\quad.
\ee
The corresponding Lax equations follow from the R-matrix formulation 
\cite{rsts}, and they are given by
\be
\dot L = [L,M_\pm ] \quad,\quad M_\pm = \frac12 (R \pm 1)  d\vf (L)
=\pm P_\pm L \quad,
\label{leom}
\ee
where $\vf (L)=\frac12 tr(L^2)$ is an invariant function on $\hat g$ 
and $R = P_+ - P_-$ is the R-matrix. $P_\pm$ are the projectors onto 
$\hat g_\pm$ subalgebras, which are defined as
\be
\hat g_+ = \sum_{j\ge 0} g_j \l^j \quad,\quad \hat g_- = \sum_{j< 0} g_j \l^j
\quad,
\ee
so that $\hat g = \hat g_+ \oplus \hat g_-$, and therefore
\be
M_+ = P_+ L = {\bf l} + {\bf a} \l \quad,\quad 
M_- =- P_- L = -{\bf s} \l^{-1}\quad. 
\ee
The equations (\ref{leom}) are solved by
\be
L(\l ,t) = g_\pm (\l ,t) L(\l) g_\pm^{-1}(\l,t)
\ee
where $L(\l)= L(\l,t=0)$ and $g_\pm (\l,t)$ are solutions of the following 
matrix factorization problem
\be
e^{tL(\l)} = g_+^{-1}(\l,t) g_- (\l,t) \quad,\label{mfp}
\ee
such that $g_+$ is analytic for $\l \ne \infty$ and $g_-$ is analytic for
$\l \ne 0$. This is a Riemann-Hilbert problem for $e^{tL(\l)}$ and a
curve which encircles the $\l=0$ point in the complex plain.

Solution of the factorization problem (\ref{mfp}) exists for $t$ that 
satisfies the Gohberg-Feldman bound, which is satisfied for sufficiently small
times \cite{rsts}. Otherwise one has a non-canonical factorization 
\be
e^{tL(\l)}=g_+^{-1}(\l,t)D(\l) g_- (\l,t) \quad,\label{ncfact}
\ee
where $D(\l)= diag(\l^{k_1},...,\l^{k_n})$.

\section{Algebro-geometric factorization}

One way to solve the problem (\ref{mfp}) is to use the connection with
Riemann surfaces. The spectral curve is given by
\be
\det (L(\l) - \m ) = 0 \quad,\label{sc}
\ee
and it is time-independent. Its compact model defines the corresponding
Riemann surface $\G$. Consider the line bundle $E_L (p)\in {\bf C}^n$ 
associated to the eigen-vectors of $L$
\be
L(p)v(p) = \m (p)v (p)\quad,\quad p\in \G \,.
\ee
Its time-evolution is linear on $Jac\, \G$, and it is given by 
$E_t = E\otimes F_t$, where $F_t$ is the line bundle on $\G$ determined by
the transition function $\exp t\m(p)$ with respect to the covering 
$\{U_+,U_-\}$, where $U_\pm = \{ p\, |\,\l^{\pm 1}(p)\ne \infty\}$. 
In order to solve
(\ref{mfp}), one needs to consider the holomorphic sections of the dual bundle
$E^*_L$, since the bundle $E_L$ does not have holomorphic sections. Denote
these sections as $\psi(p)$, then they satisfy
\be
L(p)\psi(p) = \m (p)\psi (p)\quad,\label{eve}
\ee 
so that $\psi_\pm (p,t)= g_\pm (\l(p),t)\psi (p)$ are eigenvectors of
$L(\l,t)$ which are regular in $U_\pm$. Hence
\be
g_\pm (\l,t)= \Psi_\pm (\l,t)\Psi^{-1}_\pm (\l,t=0) \label{gpsi}
\ee   
where $\Psi_\pm (\l,t)_{jk} = \psi^j (p_k,t)$. Hence if one knows the sections
$\psi_\pm$, one can solve the factorization problem (\ref{mfp}) as well as the
dynamics. These sections can be represented via meromorphic functions 
$\vf_\pm$ on $U_\pm$ as
\be
\vf_\pm (p,t) = {\psi_\pm (p,t) \over s_\pm (p)} \quad,\label{baf}
\ee
where $s_\pm (p)$ are sections of $E_L^*$. 
These functions are called Baker-Akheizer (BA) functions and satisfy
\be
(\vf_\pm ) \ge -D^* \label{bad}
\ee
where $D^*$ is a divisor of degree $g+n-1$, where 
$g$ is the genus of $\G$. 

Given (\ref{baf}) and (\ref{bad}) and the following properties of $\psi_\pm$
\be
{d\psi_\pm (p,t) \over dt} = -M_\pm (\l,t) \psi_\pm (p,t) \label{p1}
\ee
and 
\be 
\psi_+ (p,t) = e^{-\m(p)t} \psi_- (p,t)  \quad,\label{p2}
\ee
one can construct the BA functions in terms of the theta functions 
\cite{brs,agb,rsts}. Let $P_\infty = \sum_{j=1}^n P_j$ be the divisor of 
poles of
$\l$, let $p_0$ be a fixed point in $\G$ and let $D$ be a divisor of degree
$g$ such that the line bundle $E_L^*$ is associated to the divisor
\be
D^* = P_\infty + D - p_0 \quad.\label{dlb}
\ee

Let $\o_j$ be a set of normalized holomorphic Abelian differentials on $\G$
\be
\int_{a_j}\o_k = 2\p i \d_{jk} \quad,\quad \int_{b_j}\o_k = B_{jk} \quad,
\ee 
where $a_j$ and $b_j$, $j=1,...,g$ is a basis of homology cycles and $B$ is
the period matrix. Note that the Riemann theta function is defined on 
${\bf C}^g$ by quasi-periodicity conditions
\be
\th (z + 2\p i n) = \th (z) \quad,\quad \th (z + Bn) = 
e^{-\frac12 (Bn,n) - (z,n)}\th (z) \quad,\quad n\in {\bf Z}^g \label{thp}
\ee
and can be written as
\be
\th (z) =\sum_{n\in {\bf Z}^g} \exp\{\frac12 (Bn,n) + (z,n)\}\quad,
\label{thdef}
\ee
where $(X,Y)=\sum_{j=1}^g X_j Y_j$. The theta function can be considered as
a section on $Jac\, \G ={\bf C}^{g}/(2\p i n + Bm)$.

Let $A(p)$ be the Abel transform
\be
A(p)= \int_{p_0}^p \o \quad,\label{abt}
\ee 
and let $d\O$ be an Abelian differential of the second
kind, normalized by
\be
\int_{a_j}d\O = 0 \quad,\label{nadsc}
\ee
which is regular in $U_-$ such that $d\O -d\m$ is regular in $U_+$.
Then
\be
\vf^j_- (p,t) = \g_j (t) e^{t \O (p)} {\th (A(p) -c -Vt) \th (A(p) -c_0)
\over \th (A(p) -c_1)\th (A(p) -c_j)} \quad,
\ee
where $\O (p)= \int_{p_0}^p d\O$, and the constants $c_k$ and the function 
$\g_j (t)$ are determined from the following requirements

(a) $\vf^j_-$ has no discontinuity across the cuts

(b) $\vf^j_-$ is subordinate to the divisor $p_0 -P_j -D$

(c) $Res_{P_j}(\l^{-1}\vf^j_- ) = d_j$

\noindent where $d_j$ are time-independent constants.

Condition (a) is satisfied if $c + c_0 = c_1 + c_j$ and
$V_j = \int_{b_j}d\O$. Condition (b) is 
satisfied if the constants $c_k$ are chosen such that the 
divisor of zeros of $\th (A(p) -c_0)$ is $p_0 + \tilde D$, 
the divisor of zeros of $\th (A(p) -c_1)$ is $D + \tilde D$ and the
divisor of zeros of $\th (A(p) -c_j)$ is $P_j + \tilde D$, 
for some divisor $\tilde D$. Condition (c) serves to determine the functions
$\g_j (t)$ and it follows from (\ref{p1}) when $\l\to\infty$. The following
identity will be useful for our purposes
\be
V_j = \int_{b_j}d\O = -\sum_{k=1}^n Res_{P_k}(\m \o_k )\quad. \label{vid}
\ee

\section{The SO(2) Case}

In the case of the rotator, ${\bf a}= {\bf f}=diag(a,-a)$, 
\be
{\bf l} = \left(\begin{array}{cc}0 & l \\ -l & 0
\end{array}\right) \quad,
\ee
and
\be
{\bf s} = a\left(\begin{array}{cc}\cos 2\f & \sin 2\f \\ \sin 2\f &-\cos 2\f
\end{array}\right) \quad,\label{sm}
\ee
so that
\be
L = \left(\begin{array}{cc}a\l + a\l^{-1}\cos 2\f & l + a\l^{-1}\sin 2\f \\ 
-l + a\l^{-1}\sin 2\f & -a\l -a\l^{-1}\cos 2\f
\end{array}\right) \quad.
\ee

The spectral curve (\ref{sc}) is then given by
\be
\m^2 = a^2 \l^2 - 2E + a^2 \l^{-2} \quad,\label{scr}
\ee 
where $E=H$ is the energy. The equation (\ref{scr})
can be transformed into elliptic form
\be
w^2 = \l^4 - {2E\over a^2}\l^2 + 1 = (\l^2 - l_+^2)(\l^2 - l_-^2)\quad,
\label{esc}
\ee 
so that the spectral surface is a torus. We take the $a_1$ cycle to encircle 
the $[-l_+,-l_-]$ cut, while the $b_1$ cycle goes from the  $[-l_+,-l_-]$ cut
to the  $[l_-,l_+]$ cut, were it passes through to the second sheet
and it goes back to the $[-l_+,-l_-]$ cut. The normalized holomorphic 
Abelian differential is given by
\be
\o = {2\p i \over \ca}{d\l \over w} \quad,\quad {\ca} = \int_{a_1}{d\l \over w}
\quad,\label{nad}
\ee 
so that the period matrix is given by
\be
B = \int_{b_1}\o = 2\p i {\cb \over \ca} = 2\pi i \tau \quad. \label{pm}
\ee

The Abelian differential of the second kind $d\O$ must satisfy for $\l\to 0$
\be
d\O(p) = d\m + O(1)d\l= \left[\mp {a \over \l^2} + O(1)\right]d\l \quad,
\label{adsc}
\ee
and it is normalized by
\be
\int_{a_1} d\O = 0 \quad.
\ee
Hence the second-order poles are at $p=0^+$ and $p=0^-$ and one can 
construct $d\O$ in terms of the prime forms \cite{agb}. Alternatively,
from (\ref{adsc}) it follows that the corresponding Abelian integral $\O (p)$
behaves as
\be
\O(p) = \pm {a \over \l} + O(1) \quad,
\label{abint}
\ee
for $p\to 0^\pm$, and
hence it is a meromorphic function with simple poles at $0^\pm$, and with 
residues $\pm a$. A meromorphic function having only simple poles
can be specified by the position of the 
poles and with the values of the residues, and it can be constructed
in terms of the theta functions \cite{mumf}. This implies that 
\be
\O (p) = a\a \left[
{\th_1^{\prime} (A(p)- A(0^+))\over\th_1 (A(p)- A(0^+))}-
{\th_1^{\prime} (A(p)- A(0^-))\over\th_1 (A(p)- A(0^-))}\right]\quad,
\ee 
where $\a = 2\p i/\ca$ and $\th_1 (z)$ is the odd Jacobi theta function
(\ref{jf}) and $\th^{\prime}(z) = {d\th (z)\over dz}$.
Also note that the identity (\ref{vid}) gives
\be 
V = 2a\, {2\p i\over \ca}= 2a\a \quad.\label{vt}
\ee
 
We take $D = p_1 \ne e_i$, where $e_i$ are the end-points of the brunch-cuts,
and 
$P_\infty = P_1 + P_2 = \infty^+ + \infty^-$ so that
\bea
c_0 &=& \p i + \frac12 B \nonumber\\
c_1 &=& A(p_1) +\p i + \frac12 B \label{thck}\\
c_j &=& A(P_j) + \p i + \frac12 B \nonumber\quad,
\eea
and hence
\be
c= c_1 + c_j -c_0 = A(p_1) + A(P_j) +\p i + \frac12 B \quad.\label{thc}
\ee
By using
\be
\th (z + \p i + \frac12 B ) = e^{-B/8 - z/2} \th_1 (z) \quad,\label{jf}
\ee
where $\th_1 (z) = -i\th [\frac12 ,\frac12 ] (z)$ is the odd Jacobi theta
function, we obtain
\be
\vf^j_- (p,t) = \g_j (t) e^{t \tilde\O (p)} {\th_1 (-A(p) +A(p_1)+ A(P_j) +Vt) 
\th_1 (-A(p))
\over \th_1 (-A(p) +A(p_1))\th_1 (-A(p) +A(P_j))} \quad, \label{torex}
\ee 
and $\tilde\O(p) = \O(p) -V/2$.

The constants $\g_j (t)$ can be determined from the condition (c), and for this
one needs the expansion of (\ref{torex}) in $\z =\l^{-1}$ for $\l\to\infty$.
The following formulas will be useful
\be
A(p) = \int_{p_0}^{\infty^\pm}\o  +\int_{\infty^\pm}^p \o = 
A(P_j) - z_j
\ee
and
\be
z_j = \a_j \z + O(\z^2) = \pm{\a}\z + O(\z^2) \quad.
\ee
One also has for $p\to P_j$
\be
\O (p) = \O_j^0 + \O_j^1 \z + O(\z^2) \quad,
\ee
so that when $p\to P_j$ we obtain
\bea
\vf^j_- (p,t) &=& \g_j (t) e^{t \tilde\O_j^0} {\th_1 ( A_1 +Vt) 
\th_1 (-A_j)
\over \th_1 (-A_j +A_1)\a_j\th_1^{\prime} (0)}
\left[\z^{-1}  + \a_j  \left( {\O_j^1\over\a_j} t + 
{\th_1^{\prime} (A_1 + Vt)\over \th_1 (A_1 + Vt)}\right.\right.\nonumber\\
&+& \left.\left.{\th_1^{\prime} (-A_j)\over \th_1 (-A_j)}- 
{\th_1^{\prime} (-A_j + A_1)\over \th_1 (-A_j + A_1)}- 
{\th_1^{\prime\prime} (0)\over 2\th_1^{\prime} (0)}\right) +
O(\z)\right]
\quad, \label{atorex}
\eea 
where $A(p_k)=A_k$. Also when $p\to P^*_j$, where $p^* = (-w,\l)$, we have 
\be
\vf^j_- (p,t) = \g_j (t) e^{t \tilde\O_{j^*}^0} 
{\th_1 ( A_1 + A_j - A_j^* +Vt) \th_1 (-A_j^*)
\over \th_1 (-A_j^* +A_1)\th_1 (-A_j^* + A_j)}
\left[ 1 + O(\z) \right]
\quad. \label{atorexp}
\ee 
From (\ref{atorex}) it follows that
\be
d_j = \g_j (t) e^{t \tilde\O_j^0} {\th_1 ( A_1 +Vt) 
\th_1 (-A_j)
\over \th_1 (-A_j +A_1)\a_j\th_1^{\prime} (0)}\quad.
\ee

As far as the solutions of the equations of motion are concerned, these can be
obtained by inserting the $\l\to\infty$ expansions (\ref{atorex}) and
(\ref{atorexp}) of the BA functions into 
(\ref{p1}). Namely, for $\l\to\infty$ we have 
\be
\vf_-^j = \l\psi_-^j = \l\psi^j_0 + \psi^j_1 + O(1/\l) 
\ee
so that (\ref{p1}) gives
\be
{d\psi^j_0 \over dt} = 0 \quad,
\quad {d\psi^j_1\over dt} = {\bf s}\psi^j_0 \quad. \label{aeom}
\ee
By using
(\ref{aeom}) together with (\ref{atorex}) and (\ref{atorexp}) we get
$\psi^1_0 = (d_1 , 0)^T$, $\psi^2_0 = (0, d_2 )^T$ and 
\bea
\pm\a{d\over dt}\left[{\th_1^{\prime} (A_1 + Vt)\over\th_1 (A_1 + Vt)}\right]  
+ \O_1^j &=& \pm a \cos 2\f \label{cosid}\\
{d\over dt}\left[\g_j (t)e^{t\tilde\O_{j^*}^0}
{\th_1 (A_j - A_j^* + A_1 + Vt)\th_1(-A_j^*)\over\th_1 (A_1 - A_j^*)
\th_1 (A_j - A_j^*)}\right]
&=& a d_j \sin 2\f \quad.\label{sinid}
\eea

From (\ref{cosid}) we obtain
\be
\sin^2 \f = -\frac12 (\O_1 /a -1) - \a {V \over 2a}\,{d^2 \over dz^2} 
\log\th_1 (z) \quad, \label{sinsq}
\ee
where $\O_1 = \O_1^1 = -\O_1^2$ and $z= Vt + A_1$. Due to
(\ref{vid}), $V /2a= \a$ so that (\ref{sinsq}) can be written as
\be
\sin^2 \f = \con + \a^2 {\wp}(2a\a t + A_1) \quad,
\ee
where $\wp$ is the Weierstrass function. Since 
\be
{\wp} (\a^{-1} u | \a^{-1}\o_1 , \a^{-1}\o_2 ) = \a^{2} {\wp} ( u | \o_1 , 
\o_2 )\quad,
\ee
where $\o_{1,2}$ are the periods of the Weierstrass function, we get
\be
\sin^2 \f = {\wp}\left(2a t + {A_1\over \a}\right)+\con \quad.
\label{sinsqw}
\ee 

The expression (\ref{sinsqw}) can be shown to coincide with the solution of
the equations of motion obtained from the energy expression
\be
2E = {\dot\f}^2 - 2a^2 \cos 2\f \quad.\label{ee}
\ee
It follows from (\ref{ee}) that
\be
\tilde\o t = \pm \int_0^\f {d\f\over \sqrt{1 - k^2 \sin^2 \f}} \quad,
\label{eiom}\ee
where $\tilde\o = \sqrt{2(E + a^2)}$, $k^2 = 2a^2 (E + a^2 )^{-1}$ and 
$E \ge -a^2$. The elliptic integral (\ref{eiom}) implies 
\be
\sin \f = {\rm sn} (\tilde\o t , k) \quad.
\ee
However, in order to compare to (\ref{sinsqw}), we rewrite the integral
(\ref{eiom}) as
\be
2at = \pm \int_0^{\sin^2 \f}{dx \over \sqrt{4x (1-x) (k^{-2} - x)}}\quad.
\label{ewf}
\ee 
Since the Weierstrass function can be defined as
\be
u = \int_{\wp (u)}^\infty {dy \over \sqrt{4y^3 - g_2 y -g_3}}\quad,
\ee
from (\ref{ewf}) it follows that
\be
\sin^2 \vf = \wp ( 2at + u_0 )+ \con \quad,\quad Im\,u_0 \ne 0\quad,
\label{sinsqwe}
\ee 
where 
\be
u_0 = \int_0^{\infty}{dx \over \sqrt{4x (1-x) (k^{-2} - x)}}\quad.
\ee
Clearly the expression (\ref{sinsqwe}) coincides with (\ref{sinsqw}).

Now define the matrices $\Phi_\pm (\l,t)_{jk} = \vf^j_\pm (p_k,t)$,
where 
$$\vf_+ (p,t) = e^{-\m (p) t}\vf_- (p,t)\quad,$$ 
then
\be
g_\pm (\l,t)= \Phi_\pm (\l,t)\Phi^{-1}_\pm (\l,t=0) \quad. \label{mff}
\ee 
This formula follows from (\ref{gpsi}), due to relation (\ref{baf}). 

The relation (\ref{sinsqw}) can be satisfied for all times if the choice
of $p_1$ is such that $Im(A_1 /\a)\ne 0$, in accordance with (\ref{sinsqwe}).
This follows from
$\wp (u)= u^{-2} + O(u^2)$ so that the r.h.s. of (\ref{sinsqw})
diverges for
\be
2at + {A_1\over\a} = 0 \quad.\label{tc}
\ee 
Therefore if $Im(A_1/\a)\ne 0$, one obtains a physical solution.

If $p_1$ is chosen such that $Im(A_1 /\a)= 0$, then the factorization
(\ref{mfp}) will not be valid for $t$ close to $A_1/(2a\a)$. In this case one
would have a non-canonical factorization given by (\ref{ncfact}). We will
show in the next section that this case corresponds to a dynamical system
based on a non-compact extension of the $SO(2)$ group.

\section{Related problems}

Let us replace $\phi$ with $iq$ in (\ref{hrot}), and let $l=\dot q$. In this 
way we obtain a new dynamical system describing particle on a line moving in 
a potential $V(q)= -a^2 \cosh 2q$. 
The solution of the equations of motion can be obtained from the
energy 
$$2E = {\dot q}^2 - 2a^2\cosh 2q \quad,$$ 
and it is given by
\be
2at = \pm \int_{\e_k}^{\sinh^2 q}{dx \over \sqrt{4x (1+x) (k^{-2} + x)}}\quad,
\label{ncewf}
\ee 
where $\e_k =0$ for $k^2 =2a^2 /(E+ a^2 ) \ge 0$, while $\e_k = -k^{-2}$ for
$k^2 < 0$. Note that now there is no restriction $E + a^2 \ge 0$ as in the
compact case, since the potential is not bounded from bellow.
Then by using the same relations as in the compact case we get
\be
\sinh^2 q = \wp ( 2at + u_0 ) + \con \quad,\quad Im\,u_0 =0 
\quad,\label{ncsole}
\ee
since
\be
u_0 = \int_{\e_k}^{\infty}{dx \over \sqrt{4x (1+x) (k^{-2} + x)}}\quad.
\ee

This is an example of a dynamical system where $q$ diverges for finite times,
and hence one can expect that the corresponding factorization will be 
non-canonical. The replacement $\f\to iq$ corresponds to replacing the 
$SO(2)$
group with the non-compact subgroup of $SO(2,{\bf C})$. The corresponding
Lax matrix can be obtained from (\ref{lm}) by taking ${\bf f}=-{\bf a}$ and
$\f = iq$, so that
\be
L = \left(\begin{array}{cc}a\l -a \l^{-1}\cosh 2q & l - ia\l^{-1}\sinh 2q \\ 
-l - ia\l^{-1}\sinh 2q & -a\l + a\l^{-1}\cosh 2q
\end{array}\right) \quad,\label{nclm}
\ee
and $l=i\dot q$.
The spectral
curve is (\ref{scr}) with $E\to -E$, since  
$$-l^2 - 2a^2 \cos 2\f = {\dot q}^2 - 2a^2\cosh 2q = 2E\quad.$$ 
Consequently, the derivation
of the BA functions is almost the same as in the section 4, and the only
difference is in the equation (\ref{cosid}), where $a \to -a$ due to the
fact that now ${\bf f} = -{\bf a}$. This gives 
\be
\sinh^2 q = \wp ( 2at + A_1/\a )+ \con
\quad.\label{ncsolag}
\ee
The physical solutions are obtained for $Im\,A_1/\a =0$, in agreement with 
(\ref{ncsole}).
Hence the canonical factorization of $e^{tL}$ will not be possible for
$2at\to A_1/\a$.

Note that for $n\ge 3$ one obtains the spectral curves which are 
non-hyperelliptic, and in that case finding the BA functions gets difficult,
because not much is known about the non-hyperelliptic curves, 
except in some special cases, like Kowalewski top \cite{brs} (see \cite{agb}
for other examples). For example, for $n=3$ case the Lax matrix (\ref{lm}) 
gives a cubic spectral curve
\be
\m^3 + L^{(2)}\m - \det L = 0\quad,
\ee   
which is difficult to analyze.
It would be interesting to see whether some alternative method of solving
the corresponding factorization problem can be used, for example see
\cite{cs}. Preliminary study indicates that the techniques developed in 
\cite{cs} can be used to solve the $n=2$ case \cite{csp}, so that one
can expect that the $n=3$ case could be also solved.

Note that the Lax matrix (\ref{lm}) can be generalized  to
\be
L = {\bf a} \l^{m} + \sum_{j=m-1}^{-k} L_j \l^j \quad,\quad m,k \in {\bf N}
\quad,\label{lmho}
\ee
where $\bf a$ is a constant matrix while $L_j$ are the dynamical matrices 
\cite{rsts}.   
The dynamics is given by (\ref{leom}) and the corresponding factorization 
problem is given by (\ref{mfp}). In the
$n=2$ case ($L\in gl(2)$ or $sl(2)$) the BA functions can be easily 
calculated since in that
case one obtains a hyper-elliptic spectral curve 
\be
w^2 = P_{2m+2k} (\l) \quad,
\ee
where $P_{2m+2k}(\l)$ is a polynomial of order $2m+2k$. However, this system
is not physically interesting, i.e. the corresponding
Hamiltonian 
\be
H = -\frac12 Res\,tr(\l^{-1} L^2)\quad,
\ee
has no physical applications. Still, the two-by-two matrices (\ref{lmho}) 
provide non-trivial
solvable examples of the Riemann-Hilbert problem for $e^{tL}$, and one can 
compare the results to alternative methods of matrix factorization. 

\section*{Acknowledgements} 

We would like to thank J. Mourao, A. Perelomov and D. Korotkin for helpful
discussions. A.M. was supported by the grant 
PRAXIS/BCC/18981/98 from the Portugese Foundation for Science and Technology.

\end{document}